\documentclass[fleqn,usenatbib]{mnras}

\usepackage{newtxtext,newtxmath}

\usepackage[T1]{fontenc}
\usepackage{ae,aecompl}

\usepackage{graphicx}	
\usepackage{amsmath}	
\usepackage{amssymb}	
\usepackage{array}

\title[Members of globular clusters with Gaia DR2]{Extraction of globular clusters members with Gaia DR2 astrometry}

\author[I. H. Bustos Fierro \& J. H. Calder\'on]{
Iv\'an H. Bustos Fierro$^{1}$\thanks{E-mail: ivanbf@oac.unc.edu.ar}
and J. H. Calder\'on$^{1,2}\thanks{E-mail: calderonoac@gmail.com}$
\\
$^{1}$Universidad Nacional de C\'ordoba. Observatorio Astron\'omico. C\'ordoba, Argentina.\\
$^{2}$Consejo Nacional de Investigaciones Cient\'{i}ficas y T\'ecnicas (CONICET), Argentina.\\
}

\date{Accepted XXX. Received YYY; in original form ZZZ}

\pubyear{2019}

\begin{document}
\label{firstpage}
\pagerange{\pageref{firstpage}--\pageref{lastpage}}
\maketitle

\begin{abstract}
In this work we present a method to identify possible members of globular clusters using data from Gaia DR2. The method consists of two stages: the first one based on a clustering algorithm, and the second one based on the analysis of the projected spatial distribution of stars with different proper motions. In order to confirm that the clusters members extracted by the method correspond to actual globular clusters, the spatial distribution, the vector point diagram of the proper motions and the colour-magnitude diagrams are analysed. 
We apply the developed method to eight clusters: NGC 1261, NGC 3201, NGC 6139, NGC 6205, NGC 6362, NGC 6397, NGC 6712 and Palomar 13; we show the number of members extracted, the mean proper motions derived from them and finally we compare our results with other authors. 
In order to analyse the efficiency of the extraction method we perform an estimation of the completeness and the degree of contamination of the extracted members.
\end{abstract}

\begin{keywords}
astrometry -- globular clusters: general -- globular clusters: individual: NGC 1261, NGC 3201, NGC 6139, NGC 6205, NGC 6362, NGC 6397, NGC 6712, Pal 13
\end{keywords}



\section{Introduction}
\label{Intro}

Since stellar clusters are groups of gravitationally bound stars, their members are expected to occupy a rather small volume of space and to share their kinematic properties. Therefore, when space positions and velocities of stars are measured in a region of the sky containing a stellar cluster, field stars (foreground and background) will show a uniform distribution in the sky while the stars in the cluster will lie in a small region. In velocity space the field stars will show a well determined random distribution if the field is rich and a more spread and irregular distribution if the number of them is small, instead, the cluster stars always lie concentrated in a very small region. Based on these elementary characteristics, most galactic clusters have been primarily detected by an overdensity of the projected positions of stars on the sky. Several methods using also proper motions information were developed in order to separate the true members of the cluster from the field stars. Some on them are parametric methods --e.g. \cite{Vasilevskis1965}-- and others are non-parametric --e.g. \citet{Galadi1998}-- that have been applied on open and globular clusters.

The Hipparcos mission \citep{ESA1997} for the first time provided high quality 5-parameters astrometric data (celestial coordinates, proper motion and parallax) of individual stars that led to unprecedent accuracy in the determination of their distances and tangential space velocities. Although it was limited to about 120000 stars in the solar neighbourhood, some near open clusters could be studied with great detail. The successor of Hipparcos, the Gaia mission \citep{GaiaColl2016} is making a significant improvement in the quality and quantity of astrometric and photometric data for individual stars. The Second Gaia Data Release (hereafter Gaia DR2) \citet{GaiaColl2018a} allows detailed studies of the properties of galactic and extragalactic stellar systems, e.g. \citet{GaiaColl2018b}, \citet{Cantat2018a}, \citet{Cantat2018b} and \citet{Soubiran2018}.

In this work we develop and test a method to identify possible members of globular clusters from Gaia DR2 that consists of two stages: the first one is based on a clustering algorithm applied in a multi-dimensional space of physical parameters, for instance positions, proper motions and parallaxes. In most of the analysed clusters this stage alone cannot find all the members, so a second stage with a different approach is necessary in order to extract the members that have not been found up to this point. This second stage is based on the analysis of the projected spatial distribution of stars with different proper motions around the mean proper motion of the cluster. Finally, we estimate the completeness and the degree of contamination, i.e. the fraction of expected members that were extracted and the number of field stars that could have been erroneously labelled as members, respectively.

We selected the clusters in this work by the fact that we have worked, or we are working on them. Namely NGC 3201 \citep{Arellano2014}, NGC 6362 \citep{Arellano2018}, NGC 6205 \citep{Deras2019}, NGC 6397, NGC 6139, NGC 1261, NGC 6712 and Palomar 13 \citep{Yepez2019}. Their coordinates are shown in Table~\ref{table1}.

\begin{table}
	\centering
	\caption{ICRS coordinates and galactic coordinates of the selected clusters.}
	\label{table1}
	\begin{tabular}{lcccc} 
		\hline
		Cluster & RA & DE & \it{l} & \it{b}\\
		 & {\tiny(HH MM SS.SS)} & {\tiny(DD MM SS.S)} & {\tiny(deg)} & {\tiny(deg)}\\
		\hline
		NGC 1261 & 03 12 16.21 & -55 12 58.4 & 270.53871 & -52.12435\\
		NGC 3201 & 10 17 36.82 & -46 24 44.9 & 277.22878 & +08.64038\\
		NGC 6139 & 16 27 39.99 & -38 50 57.0 & 342.36464 & +06.93954\\
		NGC 6205 & 16 41 41.63 & +36 27 40.8 & 059.00947 & +40.91176\\
		NGC 6362 & 17 31 54.99 & -67 02 54.0 & 325.55452 & -17.56977\\
		NGC 6397 & 17 40 42.09 & -53 40 27.6 & 338.16501 & -11.95952\\
		NGC 6712 & 18 53 04.32 & -08 42 21.5 & 025.35410 & -04.31800\\
		Pal 13   & 23 06 44.48 & +12 46 19.2 & 087.10379 & -42.69993\\
		\hline
	\end{tabular}
\end{table}

\section{Methodology}
The extraction of the members of each cluster was performed in two stages. In this section we describe them in some detail.

\subsection{First stage}
\label{First}
The aim of this stage is to found groups of stars with similar characteristics in positions and/or motions. It was developed testing the different codes existing in Python for all the clustering algorithms provided by Scikit-Learn \citep{Pedregosa2011}, namely: K-means, affinity propagation, mean-shift, spectral clustering, hierarchical clustering, DBSCAN, gaussian mixtures and BIRCH. Each clustering algorithm was tested on different multidimensional spaces involving celestial coordinates ($\alpha, \delta$) or their gnomonic projection ($X_{t}, Y_{t}$), 
proper motions ($\mu_{\alpha*}=\mu_{\alpha}cos\delta , \mu_{\delta}$) or tangential velocities ($Vt_{\alpha},Vt_{\delta}$), and parallaxes ($\varpi$) or distances ($R$) as well as rectangular barycentric coordinates ($X, Y, Z$). 
For instance in two dimensions: [$\mu_{\alpha*},\mu_{\delta}$]; in three dimensions: [$\mu_{\alpha*},\mu_{\delta},\varpi$] and [$Vt_{\alpha},Vt_{\delta},R$]; in four dimensions: [$\alpha, \delta, \mu_{\alpha*},\mu_{\delta}$], [$X_{t}, Y_{t}, Vt_{\alpha},Vt_{\delta}$] and [$X_{t}, Y_{t}, \mu_{\alpha*},\mu_{\delta}$]; in five dimensions: [$\alpha, \delta, \mu_{\alpha*},\mu_{\delta},\varpi$], [$\alpha, \delta, Vt_{\alpha},Vt_{\delta}$, R],[ $X,Y,Z,Vt_{\alpha},Vt_{\delta}$], [$X_{t}, Y_{t}, R, Vt_{\alpha},Vt_{\delta}$].

The initial tries were performed on the fields of well-known near open and globular clusters. After these tries we noticed that when using parallaxes data, their relatively large uncertainty increased the dispersion of the entry data, consequently reducing the efficiency of the clustering algorithms. Therefore, we decided not to use parallaxes nor parameters computed from them, namely distances, rectangular barycentric coordinates and tangential velocities.
Furthermore, since the geometry of spherical coordinates causes that clusters at different declinations (or latitudes when working in galactic coordinates) be treated in different ways, we chose to use tangential plane coordinates ($X_{t}, Y_{t}$) instead of celestial coordinates. We finally decided to run the clustering algorithms in the four-dimensional space [$X_{t}, Y_{t}, \mu_{\alpha*},\mu_{\delta}$].

This first stage consists of several steps that are briefly summarized here:
\begin{itemize}
\item Download Gaia DR2 in a cone around the cluster with radius around two tidal radii from \citet{Kharchenko2013} and compute the gnomonic projection of the celestial coordinates: ($X_{t}, Y_{t}$) \citep{Astropy2013, Astropy2018}.
\item Pre-process the data ($X_{t}, Y_{t}, \mu_{\alpha*},\mu_{\delta}$) using RobustScaler in Scikit-Learn, in order to remove outliers and normalize the different dimensions. This normalization is necessary because the very different range of the values of coordinates and proper motions.
\item Run the BIRCH clustering algorithm \citep{Zhang1996} in the 4D space: [$X_{t}, Y_{t}, \mu_{\alpha*},\mu_{\delta}$].
\item Fit a 4D gaussian distribution with the candidate members of every cluster identified by the algorithm.
\item Discard candidate members that lie outside the $3\sigma$ ellipsoid and fit a new 4D gaussian distribution.
\item Retain as candidate members of the cluster only the stars within the $3\sigma$ ellipsoid of the last fitted gaussian distribution.
\item Plot the space distribution, the vector point diagram --hereafter VPD-- and the color magnitude diagram in Gaia DR2 photometric system (Gmag, BPmag $-$ RPmag)  --hereafter CMD-- with the possible members of the clusters found.
\item Verify that the space distribution, the VPD and the CMD are consistent with those of a stellar cluster. This is the only step in this stage that is still performed by a human being.
\end{itemize}

The results of the first stage of the membership determination on six globular clusters are displayed in Figure~\ref{fig01}. The candidate member stars of all the clusters in Figure~\ref{fig01} show: projected spatial positions that correspond to a spherical distribution, proper motions that are very concentrated as expected in the VPD of a globular cluster, and a CMD with the characteristic features of a globular cluster, e.g. the turn-off, the red giant branch and the horizontal branch. As can be noticed from the VPDs in Figure~\ref{fig01}, there are different situations regarding the distribution of the proper motions of the cluster members relative to that of the field stars: the proper motions of the members of NGC 3201 and NGC 6397 are clearly separated from the proper motions of the field stars, while in the remaining clusters both distributions are overlapped. Furthermore, in NGC 1261 and NGC 6205 the mean proper motion of the cluster is very close to the mean proper motion of the field stars.

This first stage was employed by \citet{Deras2019} to find the most likely members of the globular cluster NGC 6205 in order to build a clean CMD and to confirm the membership status of all known variable stars.

It is well known from studies on the density profiles of globular clusters \citep{deBoer2019} that they do not have definite borders, and diffuse stellar envelopes have been found surrounding some of them \citep{Kuzma2018}. However, it can be seen in Figure~\ref{fig01} that after the application of this stage, sharply defined borders are observed in the fields and in the VPDs. It must be noticed that they do not reflect physical features of the clusters, since they are a consequence of the $3\sigma$ cut imposed to the ellipsoid in the 4D space [$X_{t}, Y_{t}, \mu_{\alpha*},\mu_{\delta}$]. Therefore, there are members outside those artificial borders that will be extracted in the second stage with a different approach not based on any clustering algorithm.

\begin{figure*}
 \includegraphics[width=400pt]{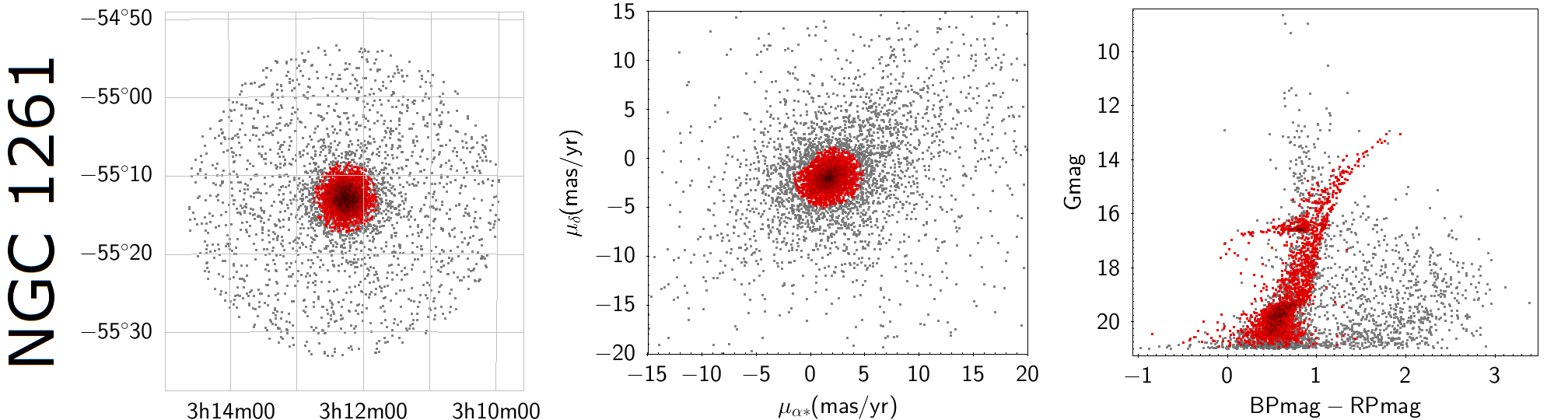}
 \includegraphics[width=400pt]{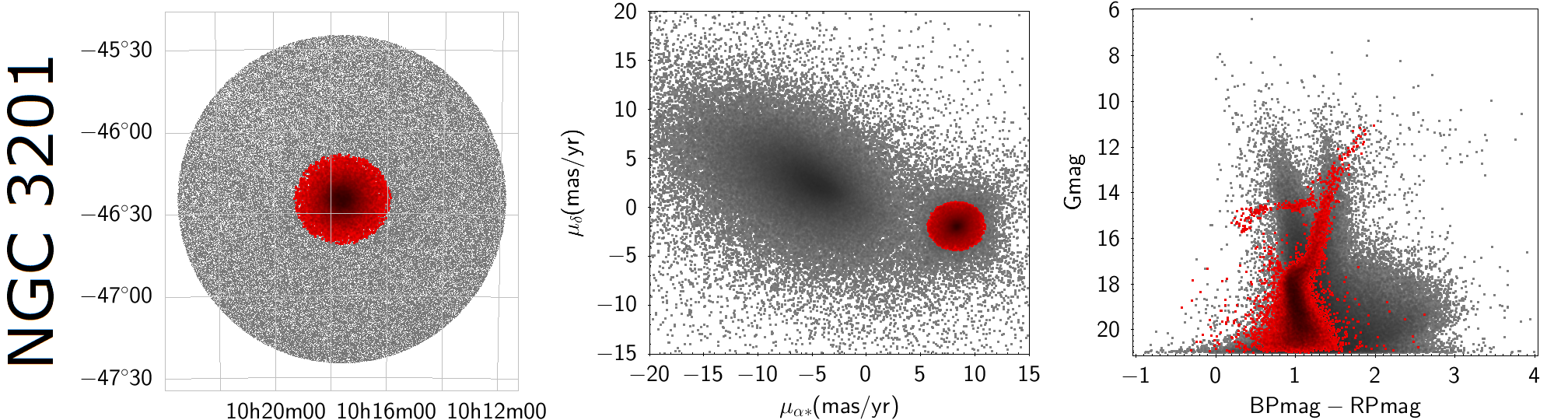}
 \includegraphics[width=400pt]{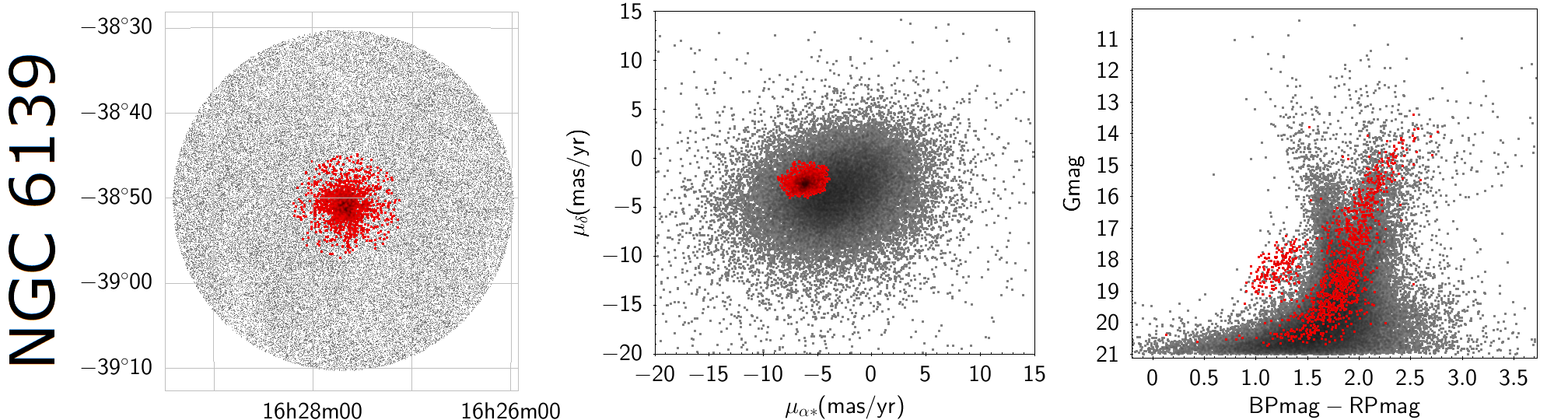}
 \includegraphics[width=400pt]{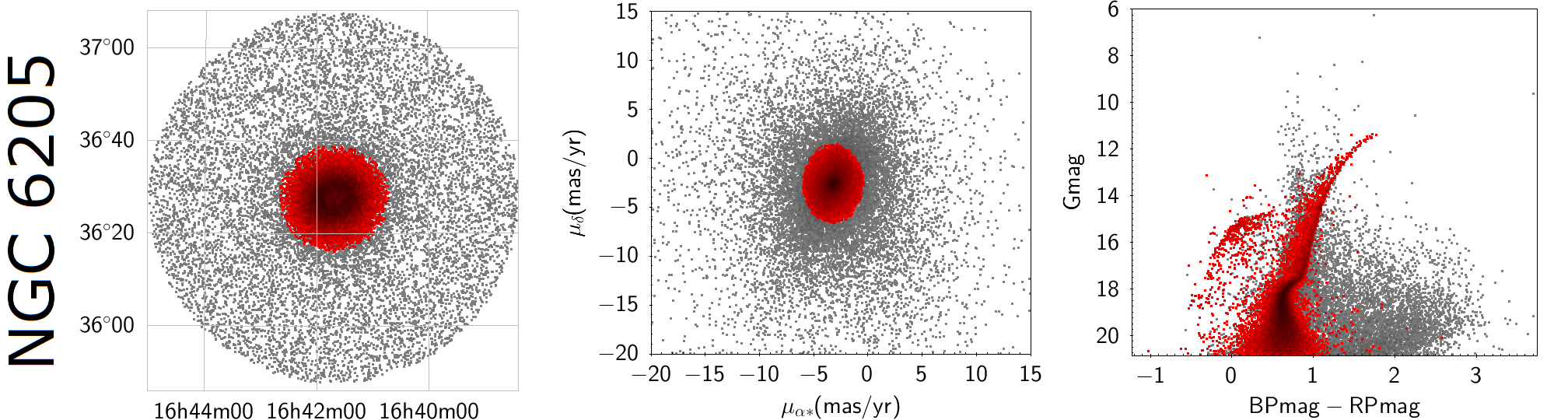}
 \includegraphics[width=400pt]{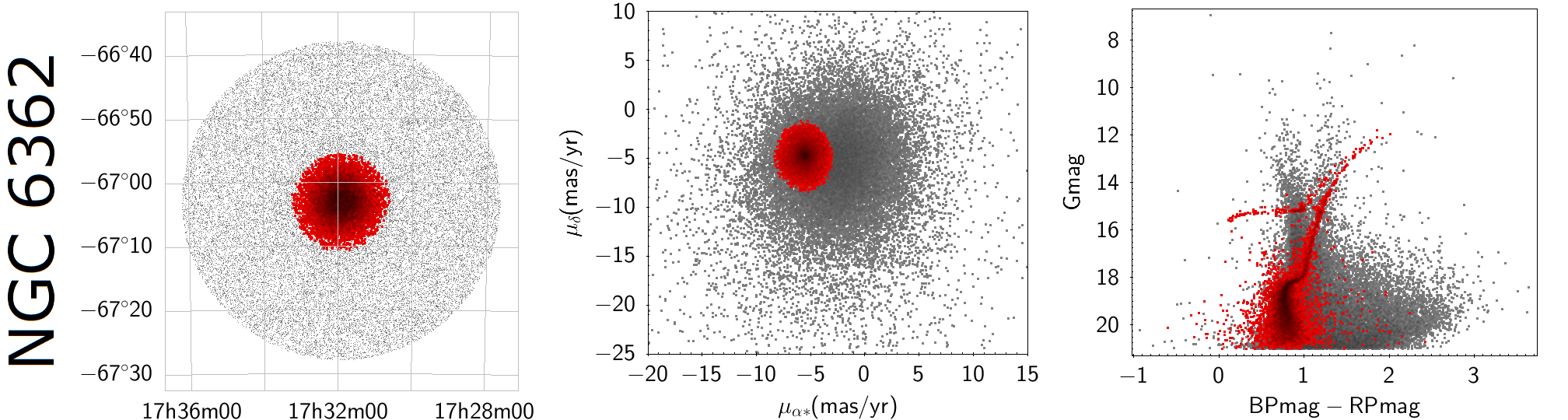}
\caption{Members extracted in the first stage (red) and remaining Gaia DR2 sample (gray) in the field of the selected clusters. Left: position in the sky, ICRS coordinates; middle: VPD; right: CMD. Notice that, as mentioned in Section~\ref{First}, the first stage only extracts a fraction of the possible members.}
 \label{fig01}
\end{figure*}

\begin{figure*}
 \includegraphics[width=400pt]{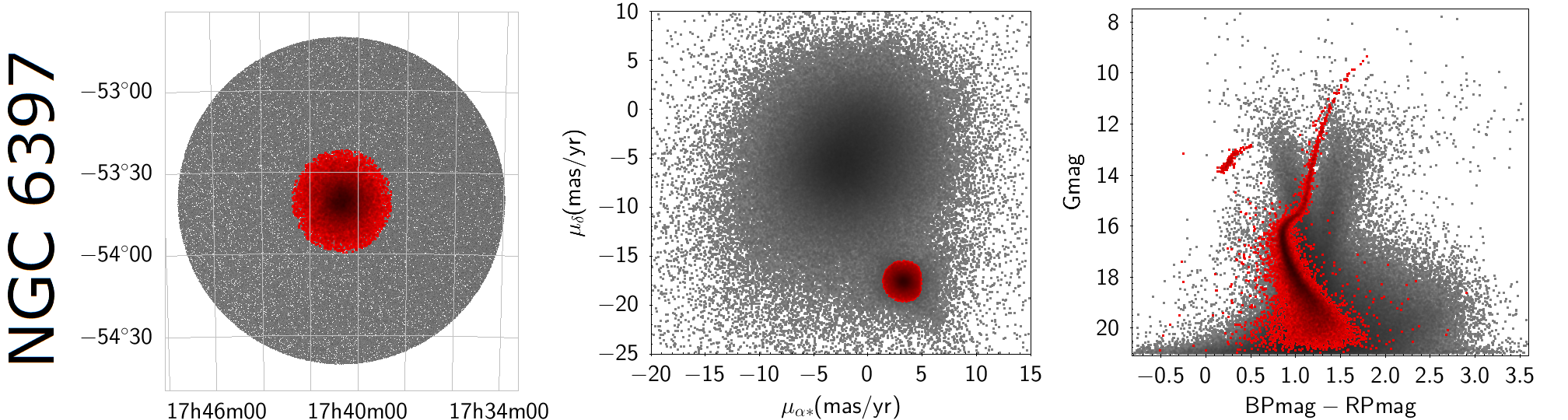}
 \includegraphics[width=400pt]{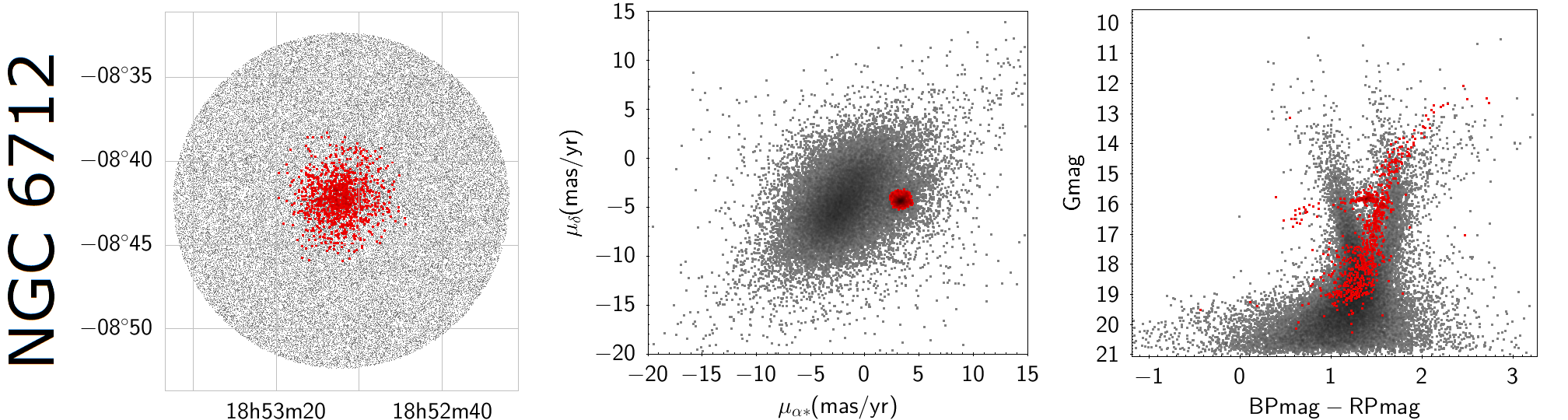}
 \includegraphics[width=400pt]{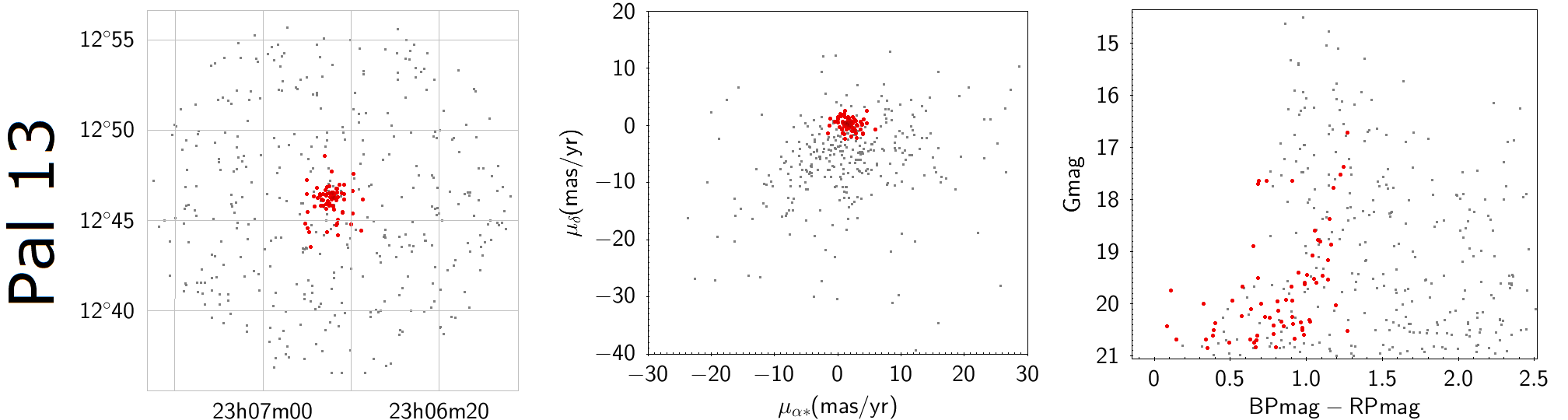}
\contcaption{Members extracted in the first stage (red) and remaining Gaia DR2 sample (gray) in the field of the selected clusters. Left: position in the sky, ICRS coordinates; middle: VPD; right: CMD.}
\end{figure*}

\subsection{Second stage}
\label{Second}
\begin{figure}
 \includegraphics[width=\columnwidth]{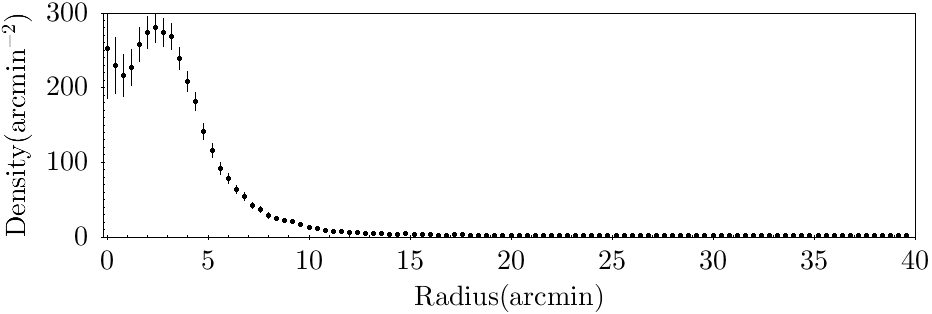}
 \includegraphics[width=\columnwidth]{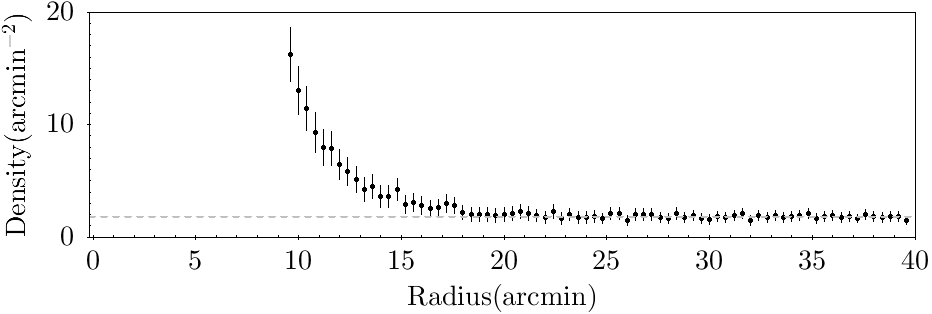}
\caption{Upper panel: Radial density projected on the sky of all the stars with measured proper motions in the field of radius $40\arcmin$ around NGC 6205, with $3\sigma$ error bars computed assuming Poisson noise in the counts (*). Lower panel: the same as upper panel with expanded vertical scale, and horizontal gray dotted line showing the adopted background density.
(*) The drop in the central region of the cluster is due to the incompleteness of Gaia DR2 proper motions in very crowded fields.
}
 \label{fig02}
\end{figure}

Any method for the determination of the members of a cluster could include some field star as members or vice versa. In order to estimate the number of missing members we computed the radial density projected on the sky of all the stars in the field with measured proper motions. As an example, in Figure~\ref{fig02} we show the radial density profile in the field of radius $40\arcmin$ around the centre of NGC 6205 that contains 40083 stars with measured proper motions. Here the background density was estimated to be 1.8 stars/arcmin$^{2}$, therefore the studied field is expected to have 9000 field stars and 31000 members of the cluster, however the first stage extracts only 23070 possible members, a number significantly less than expected. Table~\ref{table2} summarizes the results of similar analysis performed in all the fields studied in this work and the number of members extracted in the first stage.

\begin{table*}
	\centering
	\caption{Estimation of the number of members of every cluster based on the radial density profiles of stars with measured proper motion. (*) The background density is so high that the uncertainty in the number of field stars is larger than the estimated number of cluster members. (**) It is a very poor cluster, so the uncertainty in the number of field stars is larger than the estimated number of cluster members.}
	\label{table2}
	\begin{tabular}{lccccccc} 
		\hline
		Cluster & Field radius & Gaia DR2 & Stars with & Background & Expected & Expected cluster & Members extracted \\
		        & (arcmin)     & stars    & proper motion & density (arcmin$^{-2}$) & field stars & members with PM & in first stage\\
		\hline
		NGC 1261 & 20 & 9448   & 6558   & $1.02\pm0.15$   & $1280\pm190$    & $5300\pm190$ & 3024\\
		NGC 3201 & 60 & 229979 & 207158 & $15.77\pm0.35$  & $178400\pm3900$ & $28800\pm3900$ & 22200\\
		NGC 6139 & 20 & 69729  & 45384  & $34.42\pm1.04$  & $43300\pm1300$  & $2100\pm1300$ & 1462\\
		NGC 6205 & 40 & 52802  & 40083  & $1.82\pm0.15$   & $9150\pm750$    & $30900\pm750$ & 23070\\
		NGC 6362 & 25 & 54706  & 43896  & $14.32\pm0.58$  & $28100\pm1100$  & $15800\pm1100$ & 12217\\
		NGC 6397 & 60 & 472304 & 383900 & $31.55\pm0.53$  & $35700\pm6000$  & $27100\pm6000$ & 21997\\
		NGC 6712 & 10 & 60447  & 34231  & $106.20\pm3.52$ & $33400\pm1100$  & $900\pm1100$(*) & 982\\
		Pal 13   & 10 & 457    & 399    & $1.07\pm0.27$   & $340\pm90$      & $60\pm90$(**) & 69\\
		\hline
	\end{tabular}
\end{table*}

It is expected that field stars show a uniform spatial distribution, but instead in most of the fields we found that the stars not identified as members in the first stage show a residual overdensity in the same position of the cluster, and also a residual concentration in the VPD around the mean proper motion of the members (see Figure~\ref{fig03} and examples in Appendix~\ref{examples}). Notice that the stars corresponding to the peak at the mean proper motion of the members detected in the first stage lie at the outskirts of the cluster and most of them are very likely members. Actually, the stars clustered in proper motions are spread in space, while the ones clustered in spatial positions are spread in proper motions, therefore in at least one of the four dimensions they lie outside the $3\sigma$ ellipsoid in the 4D space. That is why the first stage does not detect them.

These observations and the results in Table~\ref{table2} show that there is a significant number of members that were not detected in the first stage, most of which are expected to be found in this second stage. Only in the case of Pal 13, after the first stage there are no residual overdensities neither in the field nor in the VPD, therefore the second stage was not performed.

\begin{figure}
 \includegraphics[width=\columnwidth]{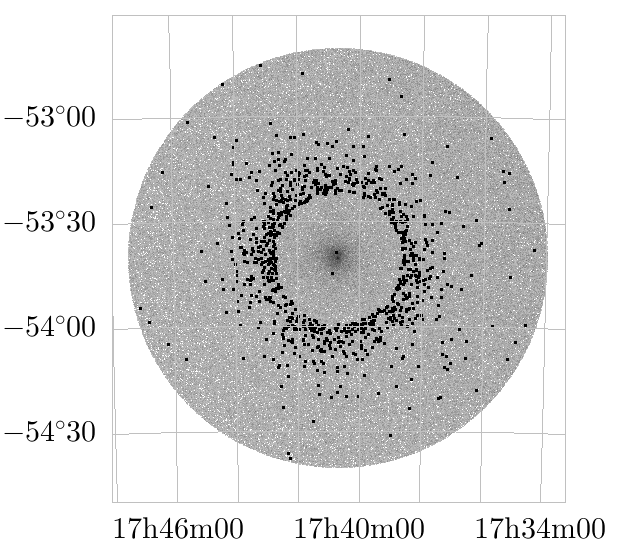}
 \includegraphics[width=\columnwidth]{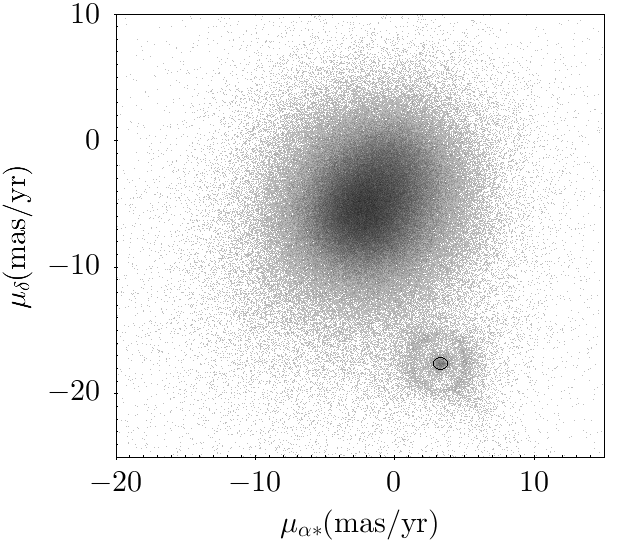}
\caption{Positions on the field (upper panel) and in the VPD (lower panel) of the stars that were not detected as members of NGC 6397 in the first stage. The stars marked with black dots in the field are the ones enclosed in the circle of radius 0.5mas/yr around the mean proper motion of the cluster members in the VPD.}
 \label{fig03}
\end{figure}

\begin{figure*}
 \includegraphics[width=400pt]{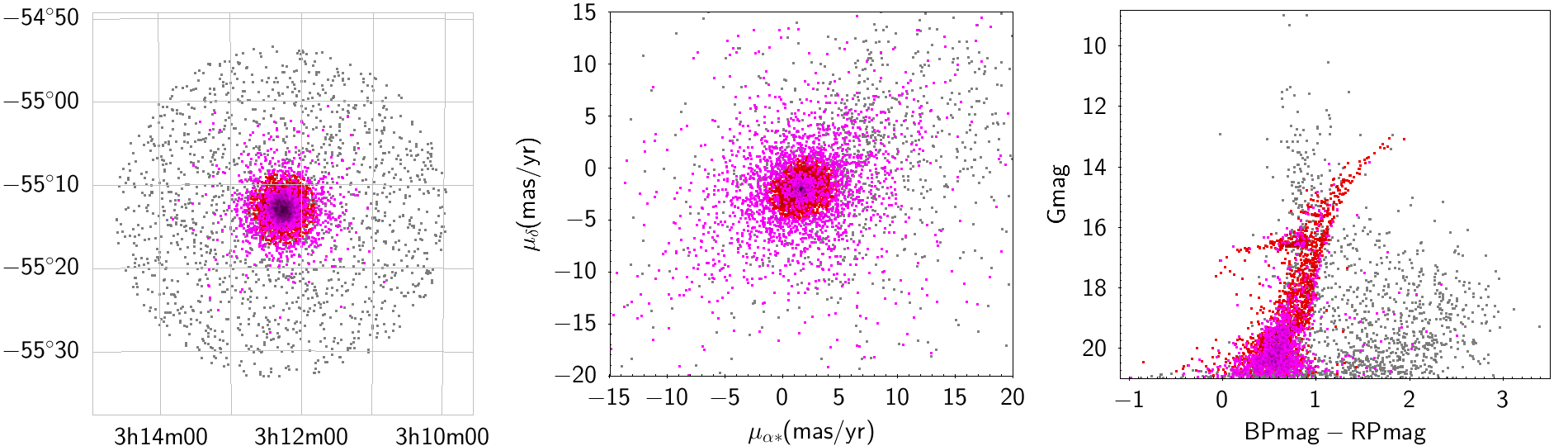}
 \includegraphics[width=400pt]{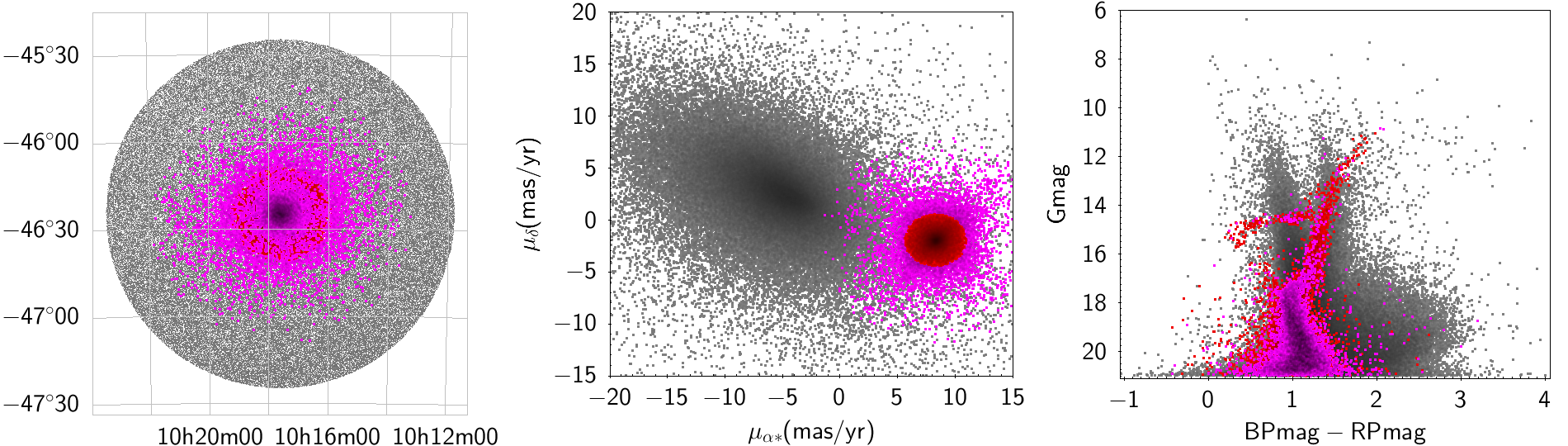}
\caption{Members extracted in the first stage (red), in the second stage (pink) and remaining Gaia DR2 sample (gray) in the field of NGC 1261 (upper panel) and NGC 3201 (lower panel). Left: position in the sky, ICRS coordinates; middle: VPD; right: CMD. Note: In all the graphics the pink dots are plotted over the red dots, except in the VPD of NGC 3201 where the pink dots are plotted under the red dots.}
 \label{fig04new}
\end{figure*}

For a better understanding of this second stage it is necessary to describe in some detail the morphological characteristics of the distribution in the VPD of the stars that were not detected as members in the first stage around the mean proper motion of the cluster. As can be seen in the lower panel of Figure~\ref{fig03}, it consists of a symmetrical radial distribution with a noticeable central peak without a definite border that is surrounded by a low-density circular area that finishes in a sharp border, after which it decays smoothly outwards. The observed symmetry in this residual distribution suggests that the best way analyse it is by means of rings of increasing radii centred in the mean proper motion of the cluster. It is remarkable to notice that this feature in VPD appears in all the analysed clusters except Pal 13, consequently one may be led to conclude that it is not related to a physical property of the cluster but a consequence of the extraction of members with the first stage of the method.

The second stage also consists of several steps that are briefly summarized here:
\begin{itemize}
\item Select rings of proper motions centred in the mean proper motion of the detected members, with increasing radii. 
\item Compute the radial density projected on the sky of all the stars within each ring of proper motions.
\item Determine for every radial density profile the mean density of the background and the radius of the central overdensity.
\item From the stars that were not classified as members in the first stage, extract as new members those that lie within each ring of proper motions in the VPD and inside the corresponding central overdensity in the field.
\end{itemize}

The outer part of all the radial density profiles is approximately constant within their errors computed assuming Poisson noise in the counts. That constant density was assumed as the background density and it was estimated as the average of the density values in that outer part.

For a better description of this stage, in Appendix~\ref{examples} we show the application on two examples representing the two different situations mentioned above regarding the proper motions: NGC 1261, where the distribution of the proper motions of the cluster and of the field stars are overlapped, and NGC 3201, where both distributions are well separated. Figure~\ref{fig04new} shows all the members of NGC 1261 and NGC 3201 that were extracted after the two stages.

\begin{table*}
	\centering
	\caption{Number of members extracted in every cluster and their mean proper motions with their uncertainties, in this work, in GC18, in BHSB19 and in Vas19. All the proper motions are expressed in mas/yr and $R_{max}$ in arcmin.}
	\label{table3}
	\begin{tabular}{ l | cccc | cccc | cc | ccc }
		\hline
		Cluster & \multicolumn{4}{c|}{This work} & \multicolumn{4}{c|}{CG18} & \multicolumn{2}{c|}{BHSB19}& \multicolumn{3}{c}{Vas19} \\
		        & Members & $\mu_{\alpha*}$ & $\mu_{\delta}$ & $R_{max}$ & Members & $\mu_{\alpha*}$ & $\mu_{\delta}$ & $R_{max}$ & $\mu_{\alpha*}$ & $\mu_{\delta}$ & Members & $\mu_{\alpha*}$ & $\mu_{\delta}$\\

		        & & $\epsilon_{\mu_{\alpha*}}$ & $\epsilon_{\mu_{\delta}}$ & & & $\epsilon_{\mu_{\alpha*}}$ & $\epsilon_{\mu_{\delta}}$ & & $\epsilon_{\mu_{\alpha*}}$ & $\epsilon_{\mu_{\delta}}$ & & $\epsilon_{\mu_{\alpha*}}$ & $\epsilon_{\mu_{\delta}}$\\
		\hline
		NGC 1261 & 5258 & 1.669 & -2.048 & 15.0 & -- & -- & -- & -- & 1.61 & -2.05 & 541 & 1.632 & -2.038\\
		         &      & 0.069 &  0.068 &      &    & -- & -- &    & 0.02 &  0.02 &     & 0.057 &  0.057\\
		NGC 3201 & 31536 & 8.298 & -1.947 & 45.0 & 19921 & 8.3344 & -1.9895 & 58.8 & 8.35 & -2.00 & 7021 & 8.324 & -1.991\\
		         &       & 0.043 &  0.043 &      &       & 0.0021 &  0.0020 &      & 0.01 &  0.01 &      & 0.044 &  0.044\\
		NGC 6139 & 2072 & -5.952 & -2.665 & 7.5 & -- & -- & -- & -- & -6.16 & -2.67 & 426 & -6.184 & -2.648\\
		         &      &  0.075 &  0.069 &     &    & -- & -- &    &  0.03 &  0.02 &     &  0.062 &  0.061\\
		NGC 6205 & 31033 & -3.176 & -2.597 & 25.0 & 15634 & -3.1762 & -2.5876 & 34.8 & -3.18 & -2.56 & 3982 & -3.164 & -2.588\\
		         &       &  0.051 &  0.052 &      &       &  0.0027 &  0.0030 &      &  0.01 &  0.01 &      &  0.047 &  0.047\\
		NGC 6362 & 16149 & -5.409 & -4.767 & 11.5 & 9169 & -5.5014 & -4.7417 & 15.6 & -5.49 & -4.76 & 2877 & -5.510 & -4.750\\
		         &       &  0.062 &  0.062 &      &      &  0.0028 &  0.0032 &      &  0.01 &  0.01 &      &  0.051 &  0.052\\
		NGC 6397 & 30526 & 3.280 & -17.573 & 33.0 & 22116 & 3.2908 & -17.5908 & 45.6 & 3.30 & -17.60 & 11406 & 3.285 & -17.621\\
		         &       & 0.039 &   0.039 &      &       & 0.0026 &   0.0025 &      & 0.01 &   0.01 &       & 0.043 &   0.043\\
		NGC 6712 & 1529 & 3.306 & -4.393 & 4.2 & -- & -- & -- & -- & 3.32 & -4.38 & 309 & 3.341 & -4.384\\
		         &      & 0.068 &  0.069 &     &    & -- & -- &    & 0.01 &  0.01 &     & 0.062 &  0.061\\
		Pal 13   & 69 & 1.805 & 0.134 & 2.8 & -- & -- & -- & -- & 1.64 & 0.25 & 34 & 1.615 & 0.142\\
		         &    & 0.341 & 0.211 &     &    & -- & -- &    & 0.09 & 0.07 &    & 0.101 & 0.089\\
		\hline
	\end{tabular}
\end{table*}

\begin{table*}
	\centering
	\caption{Sample of the table available online with the membership status of all the Gaia DR2 sources in the studied fields. Column 1: Cluster in the centre of the field; column 2: Gaia DR2 source identifier; columns 3 to 7: coordinates, proper motions and G magnitude from Gaia DR2; column 8: membership status: M1 and M2 cluster member extracted in stage 1 or 2 respectively, FS field star, UN unknown status due to proper motion not available in Gaia DR2.}
	\label{tablesample}
	\begin{tabular}{cccccccc}
		\hline
		Field &  Source & RA (ICRS) & DE (ICRS) & $\mu_{\alpha*}$ & $\mu_{\delta}$ & Gmag & Membership \\
		      &         & (deg.)  & (deg.)  & mas/yr          & mas/yr         & & status\\
		\hline
  NGC6362 & 5813078054348813952 & 263.21601570027 & -67.14175729371 & -4.504 & -1.01  & 19.9794 & FS \\
  NGC6362 & 5813078054348817280 & 263.21665026925 & -67.13299247263 & -5.541 & -4.891 & 17.8539 & M1 \\
  NGC6362 & 5813078084397967104 & 263.24072593825 & -67.13770768894 & 0.811  & -3.248 & 18.9709 & FS \\
  NGC6362 & 5813078084397982336 & 263.22608925444 & -67.13456042868 & -3.879 & -5.396 & 18.1761 & M2 \\
  NGC6362 & 5813078084397982464 & 263.22985041340 & -67.13388363600 & -5.752 & -5.313 & 19.0014 & M1 \\
  NGC6362 & 5813078084397983744 & 263.24810934919 & -67.13319244657 &        &        & 21.3399 & UN \\
		\hline
	\end{tabular}
\end{table*}

\begin{table*}
	\centering
	\caption{Estimation of the completeness of the extracted members.}
	\label{table4}
	\begin{tabular}{lccccccc}
		\hline
		Cluster & Field radius & Gaia DR2 & Background              & Expected    & Expected cluster & Total members & \% Total \\
		        & (arcmin)     & stars    & density (arcmin$^{-2}$) & field stars & members          & extracted     & members\\
		\hline
		NGC 1261 & 20 & 9448   & 1.2 & 1508 & 7940 & 5258 & 66\\
		NGC 3201 & 60 & 229979 & 17.0 & 192265 & 37714 & 31536 & 84\\
		NGC 6139 & 20 & 69729  & 51.0 & 64088 & 5641 & 2072 & 37\\
		NGC 6205 & 40 & 52802  & 2.0 & 10053 & 42749 & 31033 & 73\\
		NGC 6362 & 25 & 54706  & 16.8 & 32987 & 21719 & 16149 & 74\\
		NGC 6397 & 60 & 472304 & 38.3 & 433163 & 39141 & 30526 & 78\\
		NGC 6712 & 10 & 60447  & 180.0 & 56549 & 3898 & 1529 & 39\\
		Pal 13   & 10 & 457    & 1.2 & 380 & 77 & 69 & 90\\
		\hline
	\end{tabular}
\end{table*}

\begin{table*}
	\centering
	\caption{Estimation of the contamination of the extracted members.}
	\label{table5}
	\begin{tabular}{lccccccccc}
		\hline
		Cluster & Members   & Field stars & Contamination & Members   & Field stars & Contamination & Total   & Total       & Overall\\
		        & 1st stage & 1st stage   & 1st stage (\%)& 2nd stage & 2nd stage   & 2nd stage (\%)& members & field stars & contamination (\%)\\
		\hline
		NGC 1261 & 3024 & 32 & 1 & 2234 & 100 & 4 & 5258 & 132 & 3\\
		NGC 3201 & 22200 & 3 & 0 & 9336 & 348 & 4 & 31536 & 385 & 1\\
		NGC 6139 & 1462 & 297 & 20 & 610 & 461 & 76 & 2072 & 758 & 37\\
		NGC 6205 & 23070 & 181 & 1 & 7963 & 543 & 7 & 31033 & 724 & 2\\
		NGC 6362 & 12217 & 209 & 2 & 3932 & 982 & 25 & 16149 & 1191 & 7\\
		NGC 6397 & 21997 & 89 & 0 & 8529 & 569 & 7 & 30526 & 658 & 2\\
		NGC 6712 & 982 & 169 & 17 & 547 & 107 & 20 & 1529 & 276 & 18\\
		Pal 13   & 69 & 3 & 4 & -- & -- & -- & 69 & 3 & 4\\
		\hline
	\end{tabular}
\end{table*}

\section{Results}
In Table~\ref{table3} we show the number of members we found in Gaia DR2 for every cluster, their mean proper motions, and the radius from the cluster centre that contains all the detected members ($R_{max}$). In this work the mean proper motion of every cluster was computed as the simple average of the individual proper motions of the determined member stars, without any consideration about their errors and covariances. The uncertainties on the mean proper motions were computed considering the contributions due to random and systematic errors. The last ones were estimated using the spatial covariance of proper motions errors derived by \citet{Lindegren2018}.

For comparison we show the results of \citet{GaiaColl2018b} -GC18 hereafter-, \citet{Baumgardt2019} -BHSB19 hereafter- and \citet{Vasiliev2019} -Vas19 hereafter-. The uncertainties in our work are similar to those in Vas19 since they were calculated following an approach alike, whereas GC18 and BHSB19 do not take systematic errors into account and consequently their formal errors are smaller. Note that some clusters were not analysed by GC18, since their sample is mostly formed by clusters closer than 12kpc.

Table~\ref{tablesample} shows some rows of the table containing all the sources in Gaia DR2 in the studied fields with their membership status. The full table is available online.

\section{Efficiency of the extraction}
In order to analyse the efficiency of the extraction method we perform an estimation of the completeness --fraction of the total number of expected members that is detected-- and the degree of contamination of the extracted members --number of field stars that could have been labelled as members of the cluster--.

The developed method relies on the proper motions for the determination of the membership status of every star but there are many stars in Gaia DR2 with no measured proper motions, therefore several cluster members cannot be extracted. By assuming a constant projected background density, and that all the stars above that background are members of the cluster, it is possible to estimate the total number of members that could be found in Gaia DR2, despite their proper motions could have not been measured. \citet{Arenou2018} discuss the completeness of the Gaia DR2 catalog and they show that in very crowded regions the completeness level depends on both magnitude and local density. In the centre of globular clusters they found completeness below 75\% at Gmag\~20, and specifically in the centre of NGC 1261 and NGC 6205 almost all the stars with Gmag>19 are missing in Gaia DR2.

Performing an analysis like that mentioned in the beginning of Section~\ref{Second} but counting all the stars in the field instead of only those with measured proper motion, we obtain the estimations in Table~\ref{table4}. It must be noticed that the percentages in the last column are upper limits to the actual completeness for the most favourable cases (low density fields), since the incompleteness of the Gaia DR2 catalog itself has not been taken into account in their computations.

On the other hand, since the distributions of field stars and of cluster members are overlapped both in the sky and in proper motions, it is also expected that some field stars could have been erroneously extracted as members. Those stars are what we call contamination of the members. Table~\ref{table5} shows as field stars (columns 3 and 6) the number of field stars that are expected to be located in the same region of the sky and of the VPD that the extracted members, therefore they could have been erroneously labelled as members. The clusters more affected by contamination are those in the regions with the highest densities of field stars. Note also that in all cases the set of members extracted in the first stage is less affected by contamination than those extracted in the second stage.

\section{Conclusions}
The method developed was able to extract the members of eight globular clusters in a variety of backgrounds from a sample of Gaia DR2 stars with proper motions. A statistical analysis of members and field stars shows that between 37\% and 90\% of the expected number of members were extracted, but the number of members not extracted is mostly due to the lack of proper motions in a large fraction of stars. In the fields studied in this work, between 10\% and 43\% of the stars do not have their proper motions measured in Gaia DR2.

The mean proper motions of the clusters measured with our method are in excellent agreement with those found by other authors that make use of the same input catalogue (Gaia DR2). In addition to that our method extracted between 38\% and 98\% more members than GC18. 

The contamination of the cluster members by field stars was evaluated. In six of the eight studied clusters it was estimated to be between 1\% and 7\%. Only NGC 6139 and NGC 6712 present significantly larger estimated contaminations, presumably due to the overlap of the proper motions distributions corresponding to the cluster and the field, and the very crowded backgrounds with the highest average projected stellar densities in this study: 51 and 180 stars per arcmin$^{2}$ respectively.

The members found by means of the method presented here have been used to successfully clean the colour-magnitude diagrams of the globular clusters NGC 6205 and Pal 13, in order to derive more reliable astrophysical parameters. This method has also been used to confirm the membership of variable stars in NGC 3201, NGC 6205, NGC 6362 and Pal 13 (see references in Section~\ref{Intro}).

\section*{Acknowledgements}
This research has made use of the SIMBAD database, operated at CDS, Strasbourg, France (Wenger et al. 2000).
This research has also made use of the software TOPCAT (Taylor 2005).








\appendix
\section{Examples}
\label{examples}

This appendix illustrates the procedure of the second stage of members extraction by showing two examples. In the first one (NGC 1261) the distributions of proper motions of the cluster and of field stars are overlapped and their mean proper motions are similar. In the second one (NGC 3201) those distributions are clearly separated.

\subsection{NGC 1261}
\label{N1261}

\begin{figure}
 \includegraphics[width=200pt]{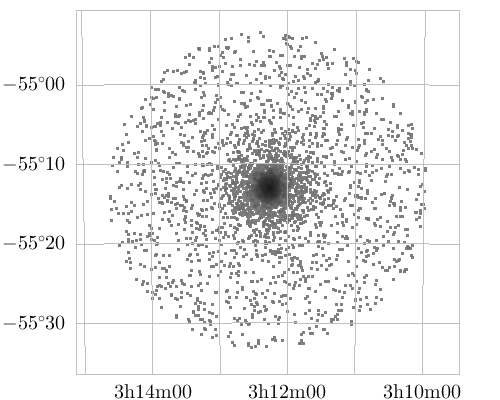}
 \includegraphics[width=200pt]{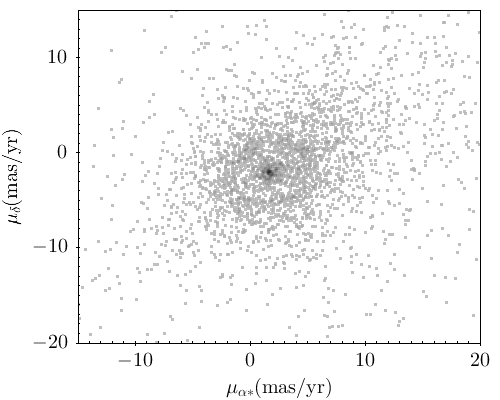}
\caption{Positions on the field (upper panel) and in the VPD (lower panel) of the stars that were not detected as members of NGC 1261 in the first stage.}
 \label{fig04}
\end{figure}

\begin{figure}
 \includegraphics[width=200pt]{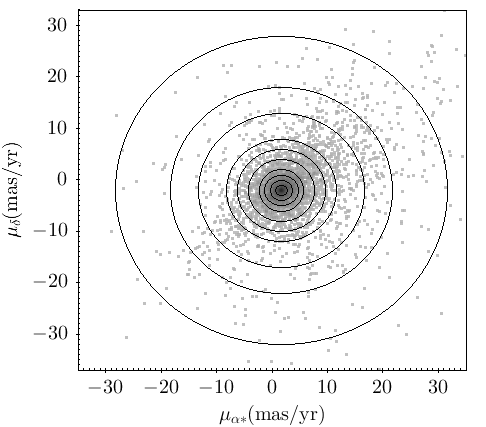}
\caption{Rings in the VPD of NGC 1261 with increasing proper motions: 0 to 1 mas/yr, 1 to 2 mas/yr, 2 to 3 mas/yr, 3 to 4 mas/yr, 4 to 6 mas/yr, 6 to 8 mas/yr, 8 to 10 mas/yr, 10 to 15 mas/yr, 15 to 20 mas/yr and 20 to 30 mas/yr.}
 \label{fig05}
\end{figure}

In Figure~\ref{fig04} we show the residual overdensity in the field and the residual concentration in the VPD around the mean proper motion of NGC 1261 after the first stage of members extraction. In this case the distribution of the proper motions of the cluster and of the field stars are overlapped, a different situation of that shown in Figure~\ref{fig03} where the distributions are clearly separated.

In the first stage we found 3024 members with mean proper motion $\mu_{\alpha*} = 1.6756 mas/yr, \mu_{\delta} = -2.0440 mas/yr$. We selected the stars in Gaia DR2 in rings of proper motions centred in this mean proper motion with increasing radii as shown in Figure~\ref{fig05}.

\begin{figure}
 \includegraphics[width=\columnwidth]{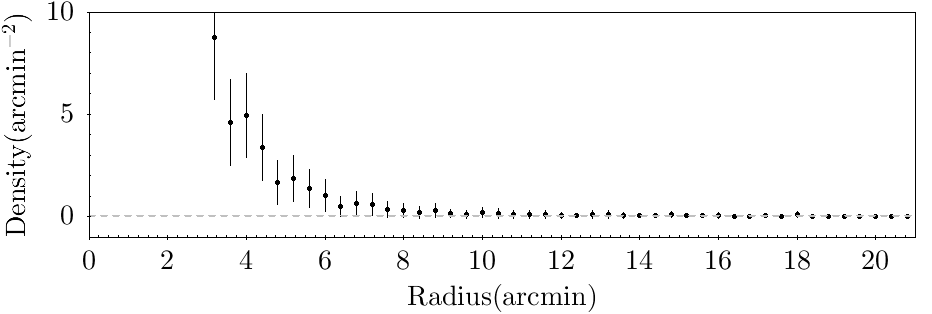}
 \includegraphics[width=\columnwidth]{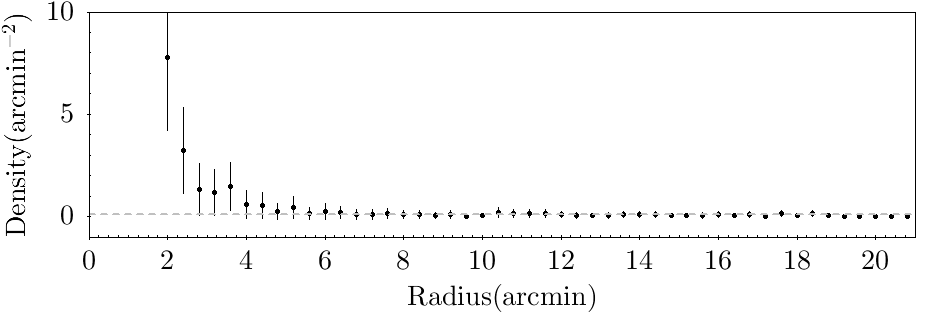}
 \includegraphics[width=\columnwidth]{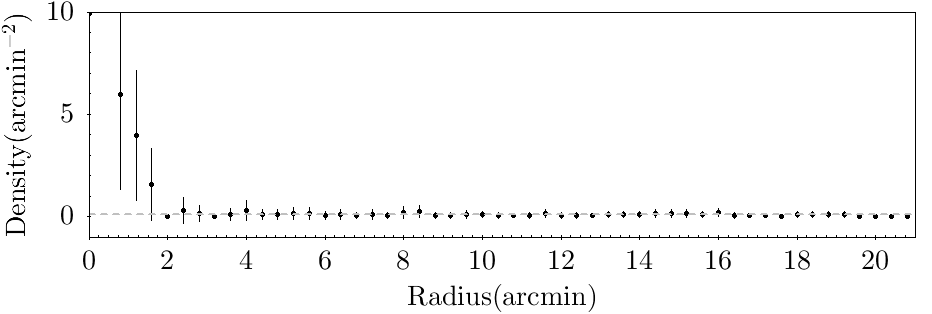}
\caption{Radial density projected on the sky of all the stars in Gaia DR2 with proper motions in three different rings, with $3\sigma$ error bars computed assuming Poisson noise in the counts, and horizontal gray dotted line showing the adopted background density in the field of NGC 1261. Upper panel: ring 0 to 1 mas/yr, middle panel: ring 3 to 4 mas/yr, lower panel: ring 20 to 30 mas/yr.}
 \label{fig06}
\end{figure}

\begin{figure*}
 \includegraphics[width=150pt]{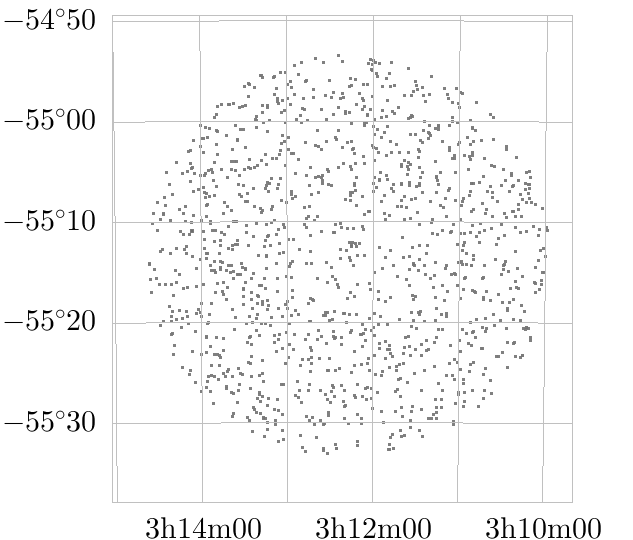}
 \includegraphics[width=150pt]{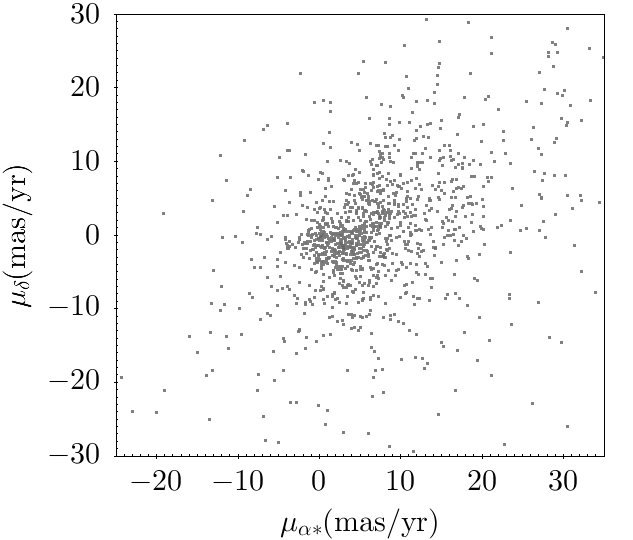}
 \includegraphics[width=150pt]{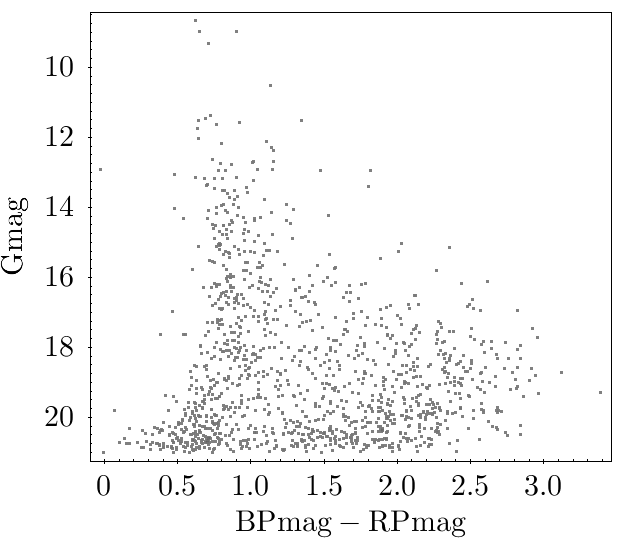}
\caption{Positions on the field (left), in the VPD (middle) and in the CMD (right) of the field stars after the two stages of extraction of NGC 1261 members.}
 \label{fig07}
\end{figure*}

We computed the radial density projected on the sky of all the Gaia DR2 stars within each ring of proper motions, some of which are shown in Figure~\ref{fig06}. It is important to notice that the stars in the innermost rings (proper motions very close to the mean proper motion of the cluster) fill a wider region on the field, while the stars in the outermost rings (larger differences in proper motions relative to the mean) are very concentrated in the central region of the cluster. This fact reflects the high velocity dispersion in that region.

\begin{figure*}
 \includegraphics[width=400pt]{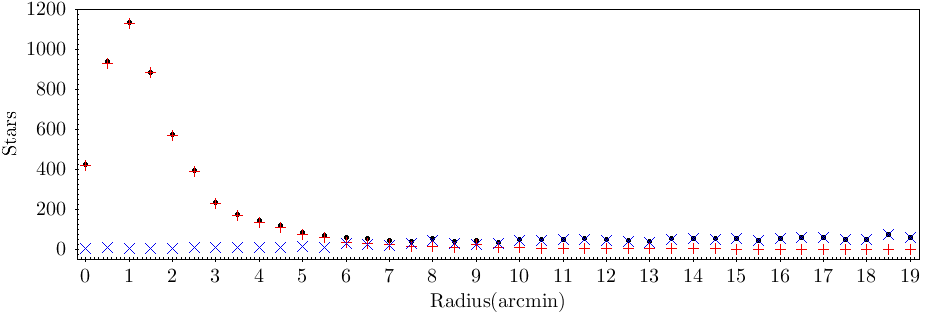}
\caption{Radial counts of field stars (blue cross), members of the cluster (red plus) and all Gaia DR2 stars with measured proper motions (black dots) in the field of NGC 1261. Notice that most of Gaia DR2 stars in this field are cluster members.}
 \label{fig08}
\end{figure*}

From the list of stars that were not classified as members in the first stage, we select those inside each ring of proper motions that in the field are lying within the radius of the central overdensity corresponding to that ring, and we add them to the list of members of the cluster. As result of this stage 2234 more members were extracted. The distribution of the resulting field stars is shown in Figure~\ref{fig07}, where there is no residual overdensity in the field nor residual concentration in the VPD, and the CMD does not show the typical features of a globular cluster.

Other way of verifying that the number of missing members is neglectable is by performing radial counts of the field stars that increases linearly with the radius if they have a constant density, as can be seen in Figure~\ref{fig08}.

\subsection{NGC 3201}
\label{N3201}

\begin{figure*}
 \includegraphics[width=\columnwidth]{FigA4a.png}
 \includegraphics[width=\columnwidth]{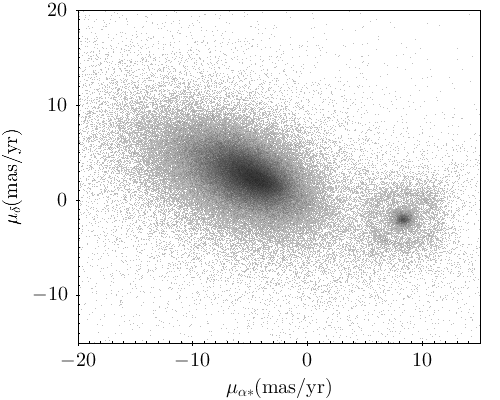}
\caption{Positions on the field (left) and in the VPD (right) of the stars that were not detected as members of NGC 3201 in the first stage.}
 \label{fig09}
\end{figure*}

\begin{figure}
 \includegraphics[width=\columnwidth]{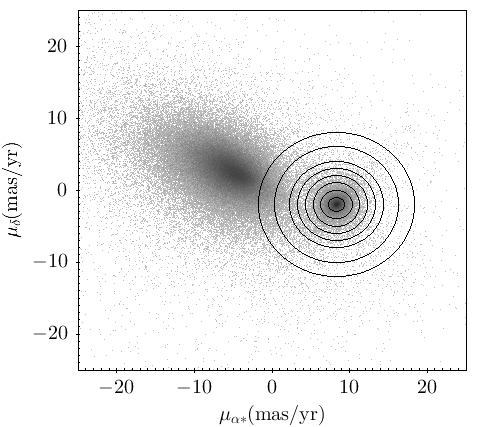}
\caption{Rings in the VPD with increasing proper motions: 0 to 1 mas/yr, 1 to 2 mas/yr, 2 to 3 mas/yr, 3 to 4 mas/yr, 4 to 5 mas/yr, 5 to 6 mas/yr, 6 to 8 mas/yr and 8 to 10 mas/yr.}
 \label{fig10}
\end{figure}

\begin{figure}
 \includegraphics[width=\columnwidth]{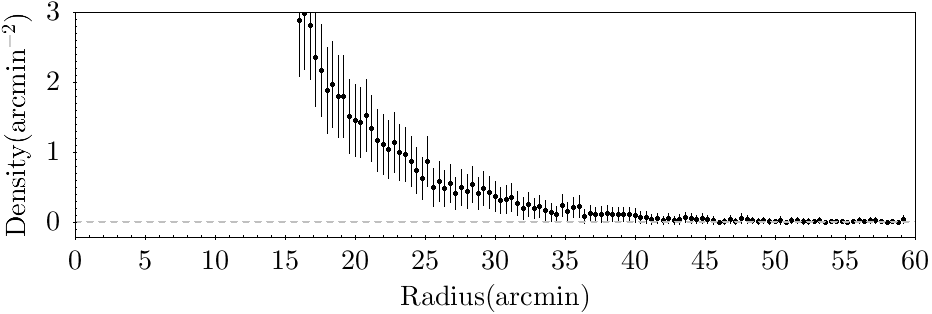}
 \includegraphics[width=\columnwidth]{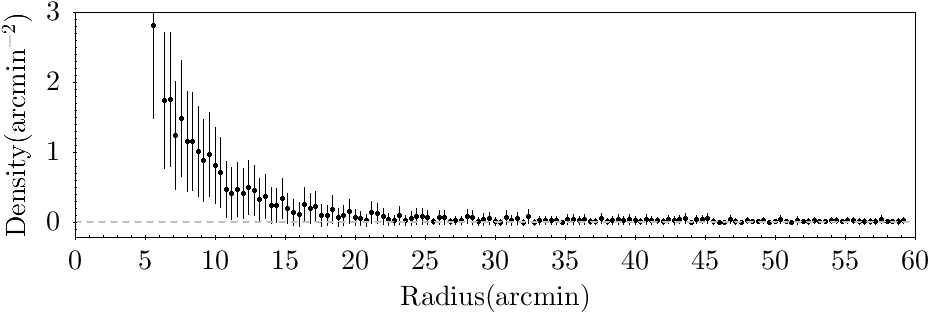}
 \includegraphics[width=\columnwidth]{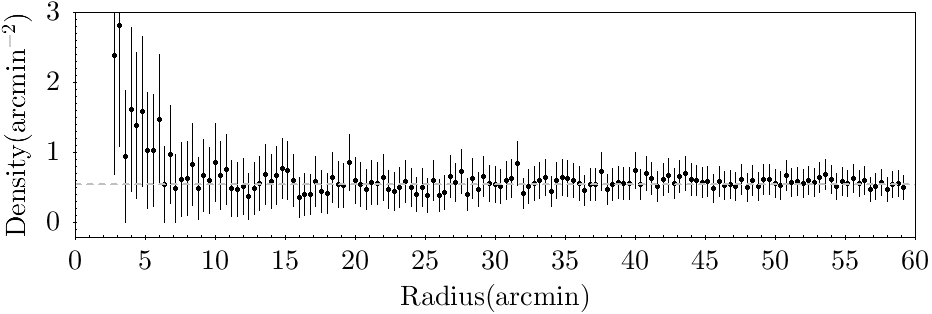}
\caption{Radial density projected on the sky of all the stars in Gaia DR2 with proper motions in three different rings, with $3\sigma$ error bars computed assuming Poisson noise in the counts, and horizontal gray dotted line showing the adopted background density in the field of NGC 3201. Upper panel: ring 0 to 1 mas/yr, middle panel: ring 3 to 4 mas/yr, lower panel: ring 8 to 10 mas/yr.}
 \label{fig11}
\end{figure}

\begin{figure*}
 \includegraphics[width=150pt]{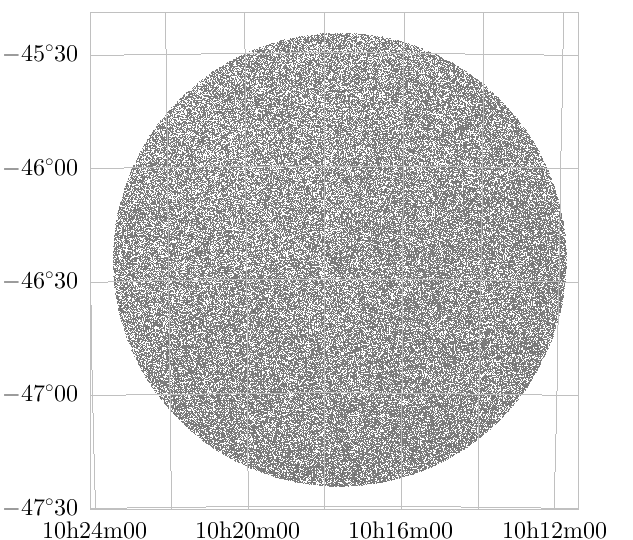}
 \includegraphics[width=150pt]{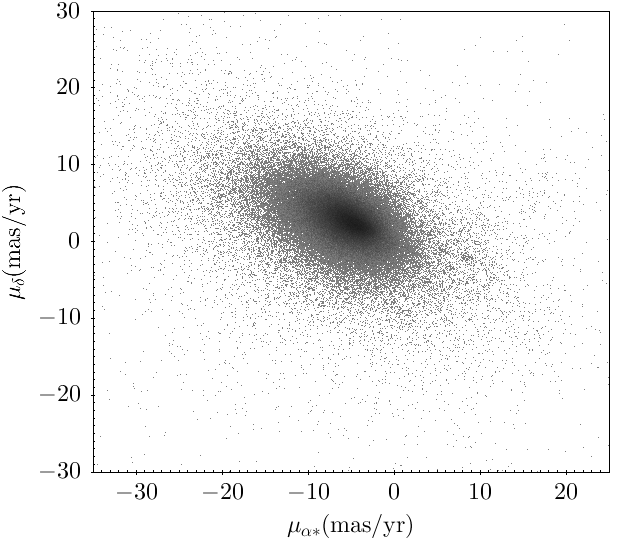}
 \includegraphics[width=150pt]{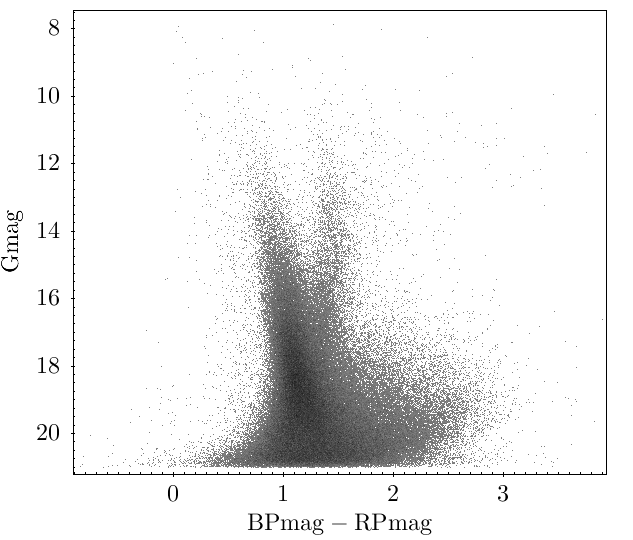}
\caption{Positions on the field (left), in the VPD (middle) and in the CMD (right) of the field stars after the two stages of extraction of NGC 3201 members.}
 \label{fig12}
\end{figure*}

\begin{figure*}
 \includegraphics[width=400pt]{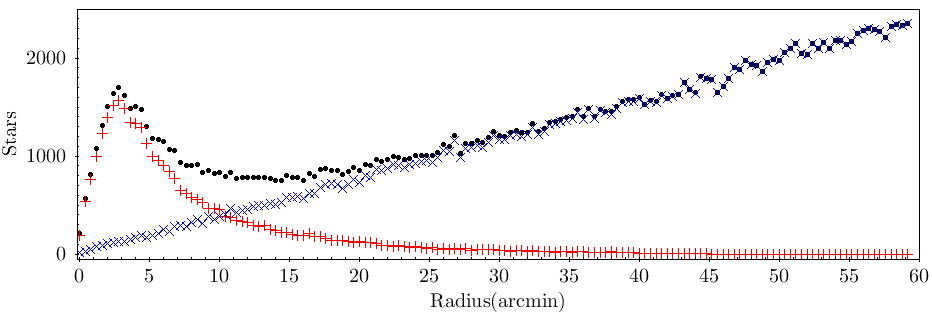}
\caption{Radial counts of field stars (blue cross), members of the cluster (red plus) and all Gaia DR2 stars with measured proper motions (black dots) in the field of NGC 3201. Notice that most of Gaia DR2 stars in this field are cluster members.}
 \label{fig13}
\end{figure*}

In Figure~\ref{fig09} we show residual overdensity and the residual concentration in the VPD around the mean proper motion of NGC 3201. In this case the distribution of the proper motions of the cluster and of the field stars are well separated, a situation like that of NGC 6397 shown in Figure~\ref{fig03}.

In the first stage we found 22200 members with mean proper motion $\mu_{\alpha*} = 8.3170 mas/yr, \mu_{\delta} = -1.9612 mas/yr$. We selected the stars in Gaia DR2 in rings of proper motions centred in this mean proper motion with increasing radii as shown in Figure~\ref{fig10}.

We computed the radial density projected on the sky of all the Gaia DR2 stars within each ring of proper motions, some of which are shown in Figure~\ref{fig11}.

From the list of stars that were not classified as members in the first stage, we select those inside each ring of proper motions that in the field are lying within the radius of the central overdensity corresponding to that ring, and we add them to the list of members of the cluster. As result of this stage 9336 more members were extracted. The distribution of the resulting field stars is shown in Figure~\ref{fig12}, where there is no residual overdensity in the field nor residual concentration in the VPD, and the CMD does not show the typical features of a globular cluster.

As in the previous section (NGC 1261) we performed the radial count of the field stars and we verified that it increases linearly with the radius as expected for a constant density (Figure~\ref{fig13}). Note that the radial count of the cluster members is smooth, while the radial count of the field stars is noisier.

\bsp	
\label{lastpage}
\end{document}